\documentclass{article}
\usepackage{graphicx}
\usepackage{amsfonts}
\usepackage{amsthm}

\newtheorem{thm}{Theorem}
\theoremstyle{definition}

\title{Open circle maps: Small hole asymptotics}
\author{Carl Dettmann\\ School of Mathematics, University of Bristol, BS8 1TW, UK}
\date{\today}

\begin{document}
\maketitle

\begin{abstract}
We consider escape from chaotic maps through a subset of phase space, the hole.
Escape rates are known to be locally constant functions of the hole position and size.
In spite of this, for the doubling map we can extend the current best result
for small holes, a linear dependence on hole size $h$, to include a smooth
$h^2\ln h$ term and explicit fractal terms to $h^2$ and higher orders,
confirmed by numerical simulations.  For more general hole locations the asymptotic
form depends on a dynamical Diophantine condition using periodic orbits ordered by
stability.
\end{abstract}

\section{Introduction}
Here we consider open dynamical systems, in which motion is considered only until the
trajectory reaches a specified subset of phase space, the ``hole(s)''.  The initial
conditions are distributed with respect to some measure, typically an invariant measure
on the phase space of the corresponding closed system (the ``escape'' problem), or
on such a measure restricted to hole(s) (``recurrence'' or ``transport'' problems).
There are also relevant questions where there are two systems connected by a hole
(the ``metastability'' problem).  There is a vast literature in mathematics and physics
on open dynamical systems and applications, some of which is mentioned in~\cite{APT,DY,D11}.

For the escape problem, we expect that for strongly chaotic systems (eg with exponential
decay of correlations), the survival probability $P(n)=\mu(M_n)$ decays exponentially
with time $n$, so that an escape rate can be defined:
\begin{equation}
\gamma=\lim_{n\to\infty}-\frac{1}{n}\ln P(n)
\end{equation}
Here $\mu$ is the measure of initial conditions (assumed to be invariant under the dynamics)
and $M_n$ the subset of the phase space $M$ that survives for at least $n$ iterations
before reaching the hole $H\subset M$. In general the limit might not exist, and we need
upper and lower escape rates, possibly infinite.  Here we restrict to maps, considering
only a discrete time $n$; similar questions apply to flows.

It is of interest to consider how the escape rate depends on the hole, in particular its
size $h=\mu(H)$.  On one hand, there have been attempts to quantify the small hole
asymptotics of the escape rate, that is, its behaviour for a sequence of holes contracting
to a point $x\in M$.  Ref.~\cite{PP} numerically 
observed the effects of periodic orbits
on the escape rate from a chaotic map with a small hole.
Ref.~\cite{BD} expresses escape rates in terms of correlation
functions, finding good numerical agreement in the case of
diamond billiard with continuous time escape rate computed
to order $h^2$; the heuristic arguments proposed
there suggest that for billiards with holes typically
considered (holes that allow particles to escape that reach
a small region of the boundary, but from any direction)
the first order result is impervious to a short periodic
orbit located in the hole, but that corrections would be
needed for the second order formulas.  Ref.~\cite{KL}
shows rigorously that
\begin{equation}\label{e:KL}
\gamma=h(1-\Lambda_x^{-1})+o(h)
\end{equation}
for a wide class of hyperbolic maps, where $\Lambda_x$ is the
stability eigenvalue of the periodic orbit starting
from $x\in M$, infinite if the orbit of $x$ is aperiodic.
This result (amongst others) was also shown independently
in a study of the open doubling map~\cite{BY}.

On the other hand it is known that the largest invariant set
that never reaches the hole, a subset of $M_\infty$, is a
locally constant function of the hole.  This is also mentioned
in~\cite{KL} but was noted much earlier~\cite{U86}.  It then
implies the escape rate, and also properties such as the
relevant Lyapunov exponents and entropy, are locally constant.
This can be seen (but was not remarked on) for the
diamond billiard in Fig.~4 of Ref.~\cite{BD}. While putting
an apparent limit on small hole asymptotic results, this effect
is also related to the very helpful observation that open
dynamics restricted to the repeller is normally a subshift
of finite type; almost always in the case of the doubling
map considered here.  For very recent results on the
existence, continuity and locally constant properties of the
escape rate for more general hyperbolic systems, see Ref.~\cite{DW}.

We note that the locally constant property does not exclude the
possibility that the escape rate is smoother as $h\to 0$ than the
single derivative required by Eq.~(\ref{e:KL}).  This paper is an
attempt to push the small hole asymptotics as far as possible in
the simplest possible dynamical system, the doubling map, with
the hope of elucidating similar effects in more general contexts.
Sec.~\ref{s:ex} discusses previous work relevant to the doubling
map.  Sec.~\ref{s:frac} considers a hole starting at zero and
uncovers an explicit, but fractal, expansion for the escape rate
to arbitrary order in $h$, including a smooth $h^2\ln h$ term as
the first contribution beyond linear. Sec.~\ref{s:H1} considers
arbitrary hole locations, finding numerical support for a general
$h^2\ln h$ term for rational locations, but rigorous arguments for
terms as slow as $h/\ln h$ for some Liouville locations, ie very
well approximable by rationals.  In more general maps, the
expansion appears to depend on an approximation theory using
periodic orbits ordered by stability.  Finally, Sec.~\ref{s:Gauss}
explores some connections with the Gauss map.    

\section{Existing results}\label{s:ex}
The doubling map $x\to 2x \pmod{1}$ has an invariant measure
$\mu$ simply given by Lebesgue, so the hole size $h$ is simply
the length of the relevant interval(s). There is a correspondence
between the binary representation of a point $x$ and
symbolic dynamics for the partition $\{[0,1/2),[1/2,1)\}$, modulo
minor details to do with the dyadic rationals (ie fractions with
denominator a power of two).

The case of a hole between 0 and $h$ has received some attention
in the literature, also including more general expanding circle
maps~\cite{J11,U86}.  In brief, given a hole size $h$, find
\begin{equation}
h_-=\sup_{x\leq h}\{2^nx>x \pmod{1},\quad \forall n\in\mathbb{N}\}
\end{equation}
This is closely related to the concept of a ``minimal
prefix''~\cite{N09}. The quantity $h_-$ is almost always a dyadic rational
$i/2^n$ for some integer $i$, and in this case $h$ lies in the interval
of constant escape rate $i/2^n\leq h\leq i/(2^n-1)$, often abbreviated
here as just ``interval.''  Note that each periodic orbit of the map
appears exactly once as the upper limit of one of these intervals.

Within an interval, the dynamics is a subshift of finite type.  If we
write the ($n$ term) binary expansion of $1-h$:
\begin{equation}
1-h_-=\sum_{k=1}^na_k2^{-k}
\end{equation}
then the leading eigenvalue of the transition matrix is
$\beta/2$ where $\beta$ is the solution of
\begin{equation}\label{e:beta}
1=\sum_{k=1}^na_k\beta^{-k}
\end{equation}
with $\beta\approx 2$.  Thus we find the escape rate
\begin{equation}
\gamma=-\ln\frac{\beta}{2}
\end{equation}
Both the local dimension of the set of irrational $h_-$ and the Hausdorff dimension
of the relevant repelling set are equal to $\ln\beta/\ln 2=1-\gamma/\ln 2$.
For irrational $h_-$ the escape rate is computed as in the rational case
(but taking $n\to\infty$) since it is continuous and monotonic.  Thus the case
of $H=[0,h]$ is of interest for $\beta$-expansions.

\section{Small hole asymptotics}\label{s:frac}
We now consider the question as to what asymptotic results are possible
in this case.  The intervals of constant escape rate
give an important constraint on the possible form of such
asympototics: $\gamma$ is constant, but $h$ has varied by
an amount proportional to $h^2$ (roughly equal to $h^2$ when $i=1$).
Thus $\gamma h^{-2}$ has jumps of unit size occurring over a short
distance relative to $h$ itself: the $O(h^2)$ term in the expansion,
if it exists, must have jump discontinuities.

For any $h$ contained in an interval, we have $\gamma(h)=\gamma(h_-)$ so
it is sufficient to consider just values $h_-$ on the left of an interval.
From the monotonicity and continuity we may furthermore calculate
$\gamma$ as a limit of rational or irrational $h_-$ as convenient.
For the following argument we use irrational $h=h_-$ and the binary
expansion of $h$ rather than the $1-h$ above:
\begin{equation}
h=2^{-J}K_0=2^{-J}\sum_{k=1}^\infty A_k2^{-k}
\end{equation}
Here, $J=(\ln K_0-\ln h)/\ln 2$ is a non-negative integer so that $K_0<1$.
There are at least two logical choices for $J$: We can set $1/2\leq K_0<1$
so that $K_0$ (and each $K_m$ below) is bounded above and below;
this is most helpful in determining the magnitude of the relevant terms.
The other choice is to set $J=0$ which simplifies the computations.
For now we need to keep $J$ arbitrary.
The other new quantities in the above expression are
$A_k=1-a_{J+k}$, with all the non-positive $A_k$ equal to
zero.  Thus the expansion for $\beta$ becomes
\begin{equation}
1=\sum_{k=1}^\infty a_k\beta^{-k}=(\beta-1)^{-1}-\sum_{k=1}^\infty A_k\beta^{-J-k}
\end{equation}

We know from~\cite{KL} that to leading order
$\gamma\approx h/2$, since the fixed point
at $x=0$ has $\Lambda=2$.  Thus we write
\begin{equation}
\beta=2-\sum_{l=1}^\infty \epsilon_l h^l
\end{equation}
where $\epsilon_1=1$, and we allow the other $\epsilon_l$ to be
functions $o(h^{-1})$ as $h\to 0$.  Substituting for $\beta$,
expanding both powers of $\beta$ in a binomial expansion,
substituting $h/K_0$ for $2^{-J}$ and equating coefficients with
equal powers of $h$, we find that
\begin{eqnarray}
\epsilon_2&=&-1+\frac{J}{2}+\frac{K_1}{2K_0}\\
\epsilon_3&=&1+\frac{3J^2}{8}+\frac{J}{8}\left(\frac{6K_1}{K_0}-11\right)+
\frac{1}{8}\left(\frac{K_2}{K_0}+\frac{2K_1^2}{K_0^2}-\frac{11 K_1}{K_0}\right)
\end{eqnarray}
and so on, where
\begin{equation}\label{e:Km}
K_m=\sum_{k=1}^\infty A_k k^m 2^{-k}
\end{equation}
consistent with the definition of $K_0$ above.  The $K_m$ for $m>0$ are
fractal functions bounded on $[1/2,1)$ as shown in Fig.~1; they
are discontinuous at dyadic rationals.

\begin{figure}
\vspace{-100pt}
\hspace{-50pt}\includegraphics[width=500pt]{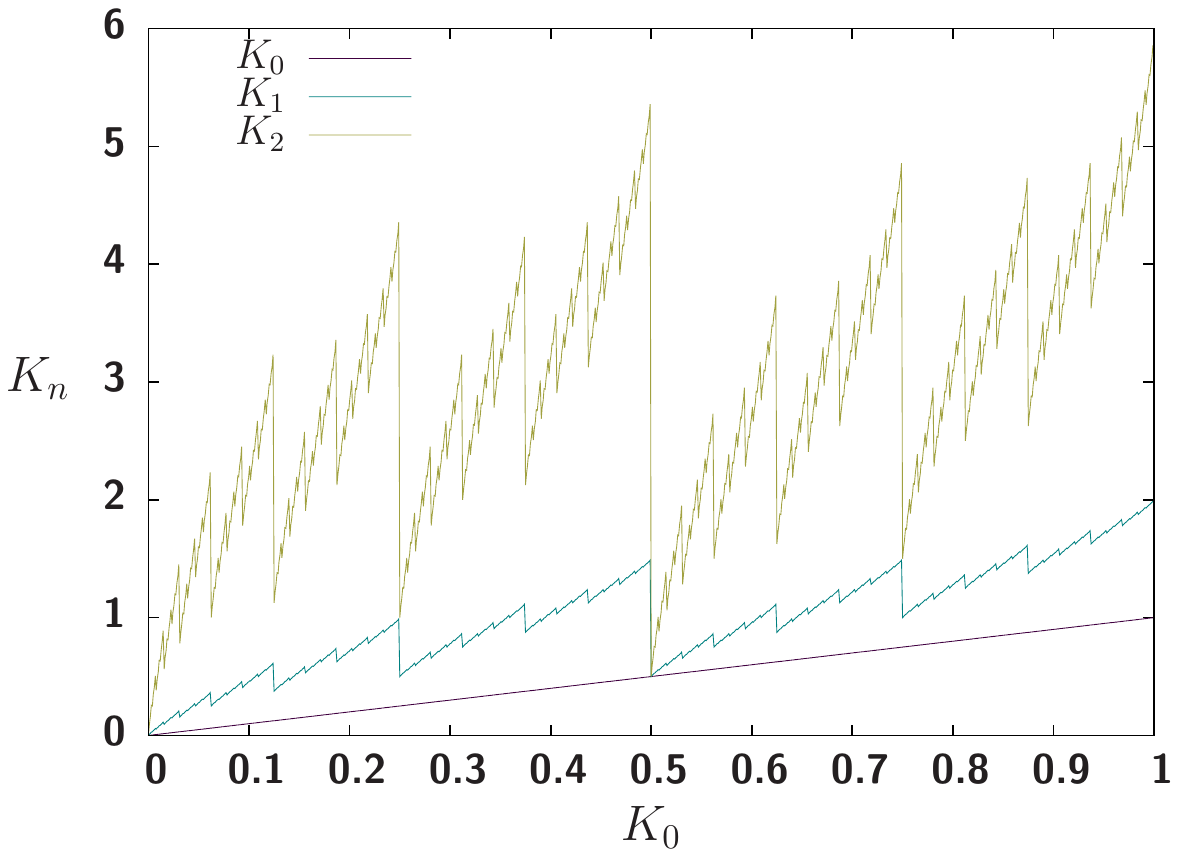}
\vspace{-400pt}
\caption{The $K_m$ functions defined in Eq.~(\protect\ref{e:Km}).}
\end{figure}

We also see that at second order the expansion involves $Jh^2$ which
gives rise to a smooth $h^2\ln h$ term. Thus we can extend
Ref.~\cite{KL} result to
\begin{equation}
\gamma=\frac{h}{2}+\frac{h^2|\ln h|}{4\ln 2}
+\frac{h^2}{4}\left(-\frac{3}{2}+\frac{\ln K_0}{\ln 2}+\frac{K_1}{K_0}\right)+\ldots
\end{equation}
for $h$ an irrational value in the fractal set not contained in any interval.

If $h$ lies in an interval $[h_-,h_+]=[i/2^n,i/(2^n-1)]$ we identify
it with the discontinuity occurring at $K_0=i/2^{(n-J)}$. The jump at
this point is $K_1^--K_1^+=2^{(1-(n-J))}$ independent of $i$, which
leads to a cancellation to second order with the discrepancy
in the first term, since $h_--h_+=i/2^{2n}$ to this order; all other
terms are smooth.  Thus, the jumps in $\gamma$ are exactly those
required by the locally constant intervals.

In what sense does this expression converge in the $h\to 0$ limit?
Considering a fixed domain of $K_0$, intervals containing $2^{-J}K_0$
decrease with increasing $J$.
Thus we expect convergence for almost all
fixed $K_0\in(1/2,1)$ as $J\to\infty$, where $h=2^{-J}K_0$, with an
error of smaller order than the last retained term.
On the other hand, there is no guarantee that for fixed $J$ the terms with
higher powers of $h$ continue to decrease. 

Recall that we did not specify $J$ exactly; we can see that the
transformation $J\to J+1$ has the effects $K_0\to K_0/2$ and
$K_1\to (K_1+K_0)/2$, leaving the above expression for
$\gamma$ unaffected.  But this also means that the coefficient of
$h^2$ is log-periodic, ie a periodic function of $\ln h$.  We can
thus write the expansion of $\gamma$ as a series involving
$h$, $h^2\ln h$ and $h^{2+2\pi i q/\ln 2}$ for $q\in\mathbb{Z}$,
however the coefficients (Fourier expansion of the $K$ functions
in the variable $\ln h$) do not appear to have a simple analytic
form.  They are also slow to converge due to the discontinuities.
We can continue analogously with $h^3$ and higher, indicating
that the Mellin transform of $\gamma$ has a semi-infinite
periodic forest of poles reminiscent of those for Julia sets~\cite{BGM}. 

Using the arbitrariness in $J$, we now set $J=0$ to yield the
simplest form of these expressions: we can substitute $h$ for
$K_0$ and evaluate the other $K$ functions at $h$, giving the
main result:
\begin{eqnarray}\label{e:main}
\gamma&=&\gamma_1+\gamma_2+\gamma_3+O(h^4\ln^3h)\\
\gamma_1&=&\frac{h}{2}\\
\gamma_2&=&\frac{2K_1h-3h^2}{8}\\
\gamma_3&=&\frac{6K_1^2h+3K_2h^2-27K_1h^2+14h^3}{48}
\end{eqnarray}
where we note that $K_m(h)\sim h|\ln^mh|/\ln^m2$ as $h\to 0$.

\begin{figure}
\vspace{-100pt}
\hspace{-50pt}\includegraphics[width=500pt]{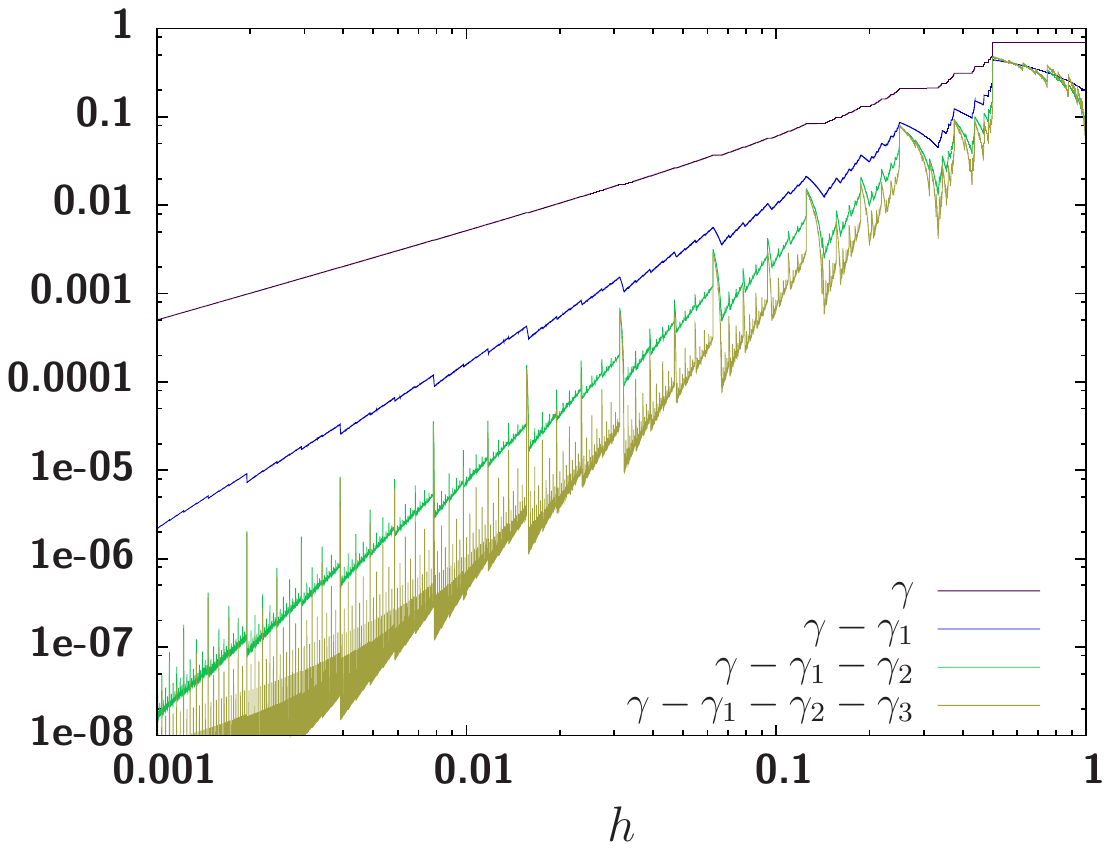}
\vspace{-400pt}
\caption{\label{f:compare} Numerical test of the small hole
expansion for the escape rate, Eq.~(\ref{e:main}).}
\end{figure}

Fig.~\ref{f:compare} shows a comparison between the escape rate
calculated directly and with the above formulas.  The direct
numerical computation uses a Markov matrix of side length $N=2^{16}$
($N=2^{19}$ for $h<0.01$)
that does not need to be stored due to its regular structure.
The leading eigenvalue corresponding to $\gamma$ is obtained by repeated
multiplication of the matrix on a vector with initially identical
nonzero entries, using long double (19 digit) precision.  The
direct escape rate is plotted, then subtracting each of the
$\gamma_i$ in turn, leading to smaller and smaller remainders.
We can see that the locally constant intervals (smooth
curves in the direct computation) become smaller for smaller $h$, but
remain as spikes at dyadic rationals.  Elsewhere, the remainder
decreases with the order of the approximation, showing good convergence
despite the fractal character of the functions involved.

Various connections may be made between the $K$ functions and
self-affine functions and sets appearing in the literature.  The
graph of $K_1$ is easily seen to be a discontinuous
self-affine set, see Ref.~\cite{P} for some relevant properties.
It shares the relation
\begin{equation}
K_1(x/2)=(K_1(x)+x)/2
\end{equation}
with the Takagi function, used to study diffusion in extended
doubling maps~\cite{KGDK}, but the similarity appears to end there,
as the Takagi function is continuous.  We can construct
a generating function for all the $K$ functions, (cf Eq.~\ref{e:beta})
\begin{equation}
G_\beta(x)=\sum_{k=1}^\infty A_k\beta^{-k}
\end{equation}
so that, at least formally, we have
\begin{equation}
K_n(x)=\left.(-\beta\partial_\beta)^nG_\beta(x)\right|_{\beta=2}
\end{equation}
The limits here are quite subtle, since $G_\beta(x)$ effectively
computes the Bernoulli convolution corresponding to $\beta$, and
it is known that the distribution of the latter is singular when
$\beta$ is Pisot, including a sequence accumulating at $\beta=2$~\cite{S};
it might be interesting to investigate these connections further.

\section{Dependence on hole position}\label{s:H1}
For holes at general locations, $H=[H_1,H_2]$ with $H_2=H_1+h$ we no longer have an explicit
formula for the characteristic polynomial (except for Markov holes; see~\cite{GDA}) so it is
not yet possible to carry out the same expansion as above. However there are a few
observations that can be made, which also apply to more general piecewise expanding maps.

\paragraph{Heuristic argument for $h^2\ln h$}
There are reasons to believe the $h^2\ln h$ term is more widely
applicable.  In a general hyperbolic map, the linear dependence of
the escape rate on the periodic orbit stability arises because that
is the extent of the ``shadow''~\cite{BP} cast by the hole on itself
under iteration around the periodic orbit.  Beyond this, we need
to study the shadow of the hole on itself along orbits
that do not remain close to the short periodic orbit (if there is one)
passing through the hole. The shortest of these are typically of
length $|\ln h|$, with stabilities of order $h^{-1}$.  This will
make a piece of the hole of size roughly $h^2$ inaccessible to
the escaping trajectories.  At the next length there will be
more periodic orbits, balanced by greater instabilities, also
yielding of order $h^2$.  This continues until double the original
length, say $2|\ln h|$ beyond which the periodic orbits are
effectively just shadowing combinations of shorter periodic orbits.
From this argument we might expect these to give a total contribution
to the escape rate of order $h^2\ln h$.  However it will depend
on the details of the periodic orbits.  We now introduce a more
systematic approach and show that reality is more subtle.

\paragraph{Relation to dynamical Diophantine properties}
Let us fix $H_1$ and take $h\to 0$.  In general we expect that the escape rate
is given (roughly) by
\begin{equation}
\gamma\approx h(1-\Lambda^{-1})
\end{equation}
where $\Lambda$ is the stability eigenvalue of the shortest periodic orbit in the interval.
The approximation arises from omitting the small $h$ limit needed for Eq.~(\ref{e:KL}).
If $H_1$ is itself aperiodic, we have $\gamma/h\to 1$ as $h\to 0$, however a short periodic
orbit $x$ lying just above $h_1$ will lead to smaller $\gamma$ when $h>x-H_1$; if there are
infinitely many such orbits (where ``short'' is relative to an appropriate function of $h$),
this may affect the limiting escape behaviour beyond linear order.

So, assume that $H_1$ is not itself periodic and consider a sequence $x_n$ of periodic points approximating $H_1$ from above.  We obtain a sequence
\begin{equation}
\gamma_n\approx h_n(1-\Lambda_n^{-1})
\end{equation}
of approximate escape rates with hole sizes $h_n=x_n-H_1$.  The correction to the linear term,
$h_n\Lambda_n^{-1}$ is of an order of magnitude relative to $h_n$ that depends on the stability
of the periodic orbit.

\paragraph{(Pre-)periodic hole locations}
For a rational $H_1=p/q$, that is, a periodic or preperiodic points for the doubling map, the
distance to a rational approximation (ie periodic point) with denominator $2^n-1$ cannot be smaller
than $1/(q(2^n-1))$, thus the size of a hole covering $H_1$ and its approximation is bounded by
a known constant multiple of $\Lambda^{-1}$ and so the correction term (making the same
convergence assumption) is order $h^2$, hence not dominating the heuristic $h^2\ln h$ term above.
A symbolic version of the same argument could also extend to more general expanding maps.
It may be possible to find explicit formulas for the characteristic polynomial in simple cases (piecewise linear map, short (pre)-periodic orbit), leading to the coefficient of $h^2\ln h$
along the lines of the previous sections, however we will confine ourselves here to a numerical
investigation of the doubling map.

We consider hole locations $H_1$ which are all the rational values with denominators in
$\{1,2,\ldots,10,15,21,31,63\}$
thus including all periodic points with periods $p\leq 6$ and also 13 preimages of
orbits with periods $p\leq 4$. The matrix size used was
$N=9999360=2^{10}\times 3^2\times 5\times 7\times 31$ and the power method
continued until a tolerance of $3\times 10^{-14}$ was reached, using 19 digit precision.  For each
location, the escape rate was computed for holes of size $h=41/N$ and $h=2^p41/N$, corresponding
to each other under the $p$ times iterated map.  The number $41$ was chosen as having no common
factor with the denominators, and roughly the smallest size consistent with the above tolerance.
Removing the known linear term from the escape rates, these two points are then used to
estimate $h^2\ln h$ and $h^2$ terms corresponding to the sequence of hole sizes found by
multiplying by powers of $2^p$.  As with the hole at zero discussed previously, the coefficient of
the $h^2\ln h$ term is found to be approximately independent of $h$, but the coefficient of $h^2$
is a fractal function of $h$.

\begin{figure}
\vspace{-100pt}
\hspace{-50pt}\includegraphics[width=500pt]{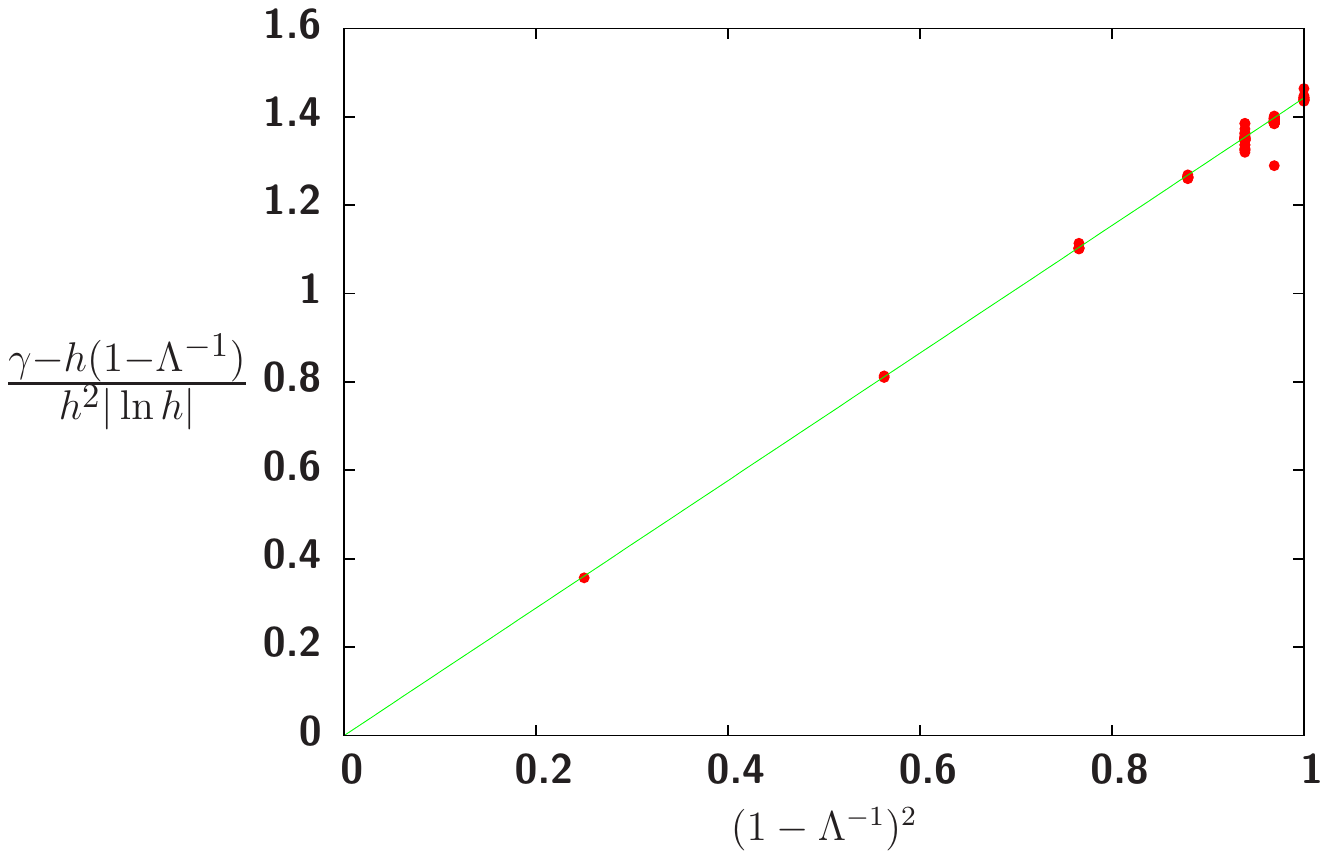}
\vspace{-400pt}
\caption{Numerical investigation of the dependence of the $h^2|\ln h|$ term for rational hole
locations; see Eq.~(\protect\ref{e:lam2}).\label{f:rat}}
\end{figure}

The results, depicted in Fig.~\ref{f:rat}, are consistent with the formula
\begin{equation}\label{e:lam2}
\gamma=(1-\Lambda^{-1})h+\frac{(1-\Lambda^{-1})^2}{\ln 2}h^2|\ln h|+O(h^2)
\end{equation}
where $\Lambda$ is taken to be infinite for preperiodic points, and the $\ln 2$ is chosen
for consistency with the previous case $H_1=0$.  The single outlier near the top right of
the figure corresponds to the point $H_1=62/63$, the period $6$ orbit most affected by the
fixed point at $1$; here convergence of the escape rate to the small hole limit is slowest,
so the extrapolation is not as accurate.  In more general cases, note that where a periodic orbit
has negative stability eigenvalue (as in the tent map for odd periods) Eq.~(\ref{e:KL}),
and hence its generalisation, need to be modified if the periodic point lies on the boundary
of the hole.

\paragraph{Typical hole locations}
Typical behaviour is given by Borel-Cantelli results, which as discussed in Ref.~\cite{A} apply to
the doubling map, and more generally.  The relevant theorem states that an infinite number
of the iterates $2^nx \pmod{1}$ of the point $x$ under the doubling map are contained in a
shrinking sequence of intervals $I_n$ for almost all $x$ if and only if the sum of the measures
(ie lengths) of the intervals diverges.
For example, the intervals $[H_1-(n\ln^k n)^{-1},H_1]$ contain infinitely
many $y_n=2^nH_1\pmod{1}$ for almost every $H_1$ if and only if $k\leq 1$.  Each such point leads to
a periodic point $x_n=H_1+(h_1-y_n)/(2^n-1)$ lying in $[H_1,H_1+((2^n-1)n\ln^k n)^{-1}]$.  Thus we expect
an infinite sequence of holes $h_n$ of size as low as $((2^n-1)n\ln n)^{-1}$ with escape rate
\begin{equation}
\gamma_n\approx h_n(1-2^{-n})\approx h_n(1-h_n|\ln h_n|\ln |\ln h_n|)
\end{equation}
This argument is not rigorous as it depends crucially on the convergence rate of Eq.~(\ref{e:KL})
but suggests that for typical $H_1$ a bound on the correction term might need to be slightly
greater than $|h^2\ln h|$.

\paragraph{The well approximable case}
In contrast to the above heuristic and numerical arguments,
the following negative result is rigorous:

\begin{thm}
Given any continuous strictly increasing function $g$ defined on $[0,1/3]$ with $g(0)=0$ and
$g(1/3)>1/4$, there is an irrational (ie aperiodic) $H_1$ such that the doubling map with
hole $[H_1,H_1+h]$ has escape rate $\gamma(h)$ so that $\gamma(h)/h-1$ approaches zero
but is not $O(g(h))$ as $h\to 0$.
\end{thm}

{\em Proof:} Consider
\begin{equation}
x_K=1-\sum_{k=1}^K\frac{1}{2^{n_k}-1}
\end{equation}
for a strictly increasing sequence $n_k$ with $n_1=2$.  Clearly all the $x_K$ lie in the unit
interval.  Then, we define
\begin{equation}
H_1=\lim_{K\to\infty} x_K
\end{equation}
and construct the sequence $n_k$ so that $H_1$ satisfies the required property.

First, assume that $n_k$ divides $n_{k+1}$ for all $k$.  Then, $2^{n_k}-1$ divides $2^{n_{k+1}}-1$
and so $x_K$ is a rational with denominator $2^{n_K}-1$ (or a factor) and hence a periodic orbit
of length $n_K$ (or a factor).  Furthermore, since the sequence is strictly increasing,
$n_{k+1}\geq 2n_k$ which implies that the limit $H_1$ is irrational as required.

There are two further lower bounds on the growth rate of $n_k$.  One comes from the limit in
Eq.~(\ref{e:KL}): There exists a function $H(\epsilon)$ (implicitly depending on $x_K$) such that
for all $h<H(\epsilon)$ we have
\begin{equation}
|\gamma(h)/h-(1-2^{-n_k})|<\epsilon
\end{equation}
considering the hole $[x_K-h,x_K]$ shrinking to the periodic orbit $x_K$.  We choose
$\epsilon=2^{-2n_k}$ to ensure that $\gamma(h)/h$ is sufficiently close to that determined
from the periodic orbit.  This requires $n_{k+1}>-\log_2 H(2^{-2n_k})+1$, where the final
term ensures that the hole can extend downward beyond $x_{K+1}$ to $h_1$.

Finally, we need to ensure that the correction $2^{-n_k}$ is greater than $g(h)$ for the hole
$[h_1,x_K]$.  This is accomplished by requiring $n_{k+1}>-\log_2 G^{-1}(2^{-n_k})+1$ where
$G^{-1}$ is the inverse of $G$, an arbitrarily chosen monotonic function with the same
conditions as $g$ but with $G(h)\neq O(g(h))$ as $h\to 0$. Thus any sequence satisfying
$n_k$ divides $n_{k+1}$ and
\begin{equation}
n_{k+1}>\max[-\log_2H(2^{-2n_k})+1,-\log_2 G^{-1}(2^{-n_k})+1]
\end{equation}
satisfies the requirements of the theorem. \qed 

Examples: For $g(h)=h^{\alpha}$ with $0<\alpha<1$ we need at least $n_{k+1}>n_k/\alpha$.  For
$g(h)=-1/\ln h$ (so that $\gamma(h)=h+O(h/\ln h)$) we need at least $n_{k+1}>C2^{n_k}$, which
has extremely rapid growth leading to $H_1$ as a Liouville number.

Remarks: The main idea of the above proof is to find aperiodic points well approximated by
periodic orbits of the map, in particular such that the separation is much smaller than the
inverse stability of the periodic orbit, $\Lambda^{-1}$.  Approximation by periodic orbits has
also appeared in different contexts, for example in obtaining the multifractal properties of
Lyapunov exponents~\cite{PW}.  Ordering periodic orbits by stability is a more generally useful
technique for systems with widely varying expansion rates~\cite{DM}.

\section{Epilogue: The Gauss map}\label{s:Gauss}
Finally, there is an intriguing connection with $h^2\ln h$ in the literature, namely for
the Gauss map $x\to\{1/x\}$ with invariant measure $d\mu=(\ln 2(1+x))^{-1}dx$.
The natural symbolic dynamics $x\to\lfloor 1/x\rfloor$ gives the partial
quotients $a_k$ of the continued fraction expansion $x=[0;a_1a_2\ldots]$.
The periodic orbits are quadratic irrationals. Thus, a hole $H=[0,1/(n+1)]$
corresponds to escape when $a_k> n$.  The hole size is
$h=\int_H d\mu=\ln(1+(n+1)^{-1})/\ln 2$.  Hensley~\cite{H92}
analyses this problem using the transfer operator
\begin{equation}
L_{s,n}f(t)=\sum_{k=1}^n\frac{f(1/(k+t))}{(k+t)^s}
\end{equation}
giving its leading eigenvalue $\lambda(s,n)$ in his Eq.~(7.9) as
\begin{equation}
\lambda\left(2-\frac{\theta}{n},n\right)=1+\left(\frac{\pi^2\theta-12}{12\ln 2}\right)\frac{1}{n}
-\frac{\theta\ln n}{n^2\ln 2}+O(n^{-2})
\end{equation}
Setting this equal to unity, he finds that the value of $\theta$, and hence the
Hausdorff dimension of the invariant set, to contain a term proportional to
$\ln n/n^2$, ie also $h^2\ln h$.  The escape rate, given by $-\ln\lambda(2,n)$,
has however no such term!

For more general hole positions, the previous discussion suggests that the
asymptotics depends on the approximation of the hole position using periodic orbits
ordered by stability.  The Gauss map has infinitely many branches, however there
are only finitely many periodic orbits with a given stability bound.  It would
thus be interesting to analyse the escape rate for general hole positions using
approximation by quadratic irrationals.

\section*{Acknowledgements}
The author gratefully acknowledges helpful discussions with Mark
Demers during his visit to Bristol in June 2011, which was partly
funded by a London Mathematical Society scheme 2 grant, and also
discussions with Eduardo Altmann, Andrew Ferguson, Orestis Georgiou,
Thomas Jordan, Rainer Klages and Georgie Knight.

\end{document}